# Epitaxial EuO Thin Films on GaAs

A. G. Swartz[1], J. Ciraldo[1], J. J. I. Wong[1], Yan Li[1], Wei Han[1], Tao Lin[1], S. Mack[2], J. Shi[1], D. D. Awschalom[2], and R. K. Kawakami[1*]

[1]Department of Physics and Astronomy, University of California, Riverside, CA 92521

[2]Center for Spintronics and Quantum Computation, University of California, Santa Barbara, CA 93106

**Abstract:**

We demonstrate the epitaxial growth of EuO on GaAs by reactive molecular beam epitaxy. Thin films are grown in an adsorption-controlled regime with the aid of an MgO diffusion barrier. Despite the large lattice mismatch, it is shown that EuO grows well on MgO(001) with excellent magnetic properties. Epitaxy on GaAs is cube-on-cube and longitudinal magneto-optic Kerr effect measurements demonstrate a large Kerr rotation of 0.57 , a significant remanent magnetization, and a Curie temperature of 69 K.

Stoichiometric EuO is an attractive material for spintronics because it is a ferromagnetic insulator with a large exchange splitting of its conduction band [1], as well as having the largest magneto-optic response of any oxide [2], and large magnetic moment of 7 Bohr magnetons ($\mu_B$) per Eu atom [3]. When used as a tunnel barrier, the EuO is an effective spin filter [1] due to its spin dependent barrier height. EuO's insulating nature enables its use as a gate dielectric for which it is predicted to generate a gate tunable exchange field for spin manipulation [4]. Historically, the growth of stoichiometric EuO has been notoriously difficult to achieve. For reactive molecular beam epitaxy (MBE), the growth requires fine-tuning of the Eu and $O_2$ fluxes because a low oxygen flux results in the formation of Eu-rich EuO ($EuO_{1-\delta}$) [5,6]. On the other hand, a high oxygen flux leads to the formation of non-magnetic $Eu_2O_3$, which is more thermodynamically favorable [7]. While $EuO_{1-\delta}$ is interesting for its metal-insulator transition and colossal magnetoresistance, its ferromagnetic phase is metallic and therefore does not possess the unique properties of a ferromagnetic insulator.

Recently, the high quality and reliable epitaxial growth of stoichiometric EuO on oxide substrates has been accomplished using a high temperature, adsorption-controlled growth mode [8-11]. In this regime, the Eu flux is set to be much higher than the $O_2$ flux and a high substrate temperature (~450 °C) is maintained to re-evaporate any excess Eu. Only Eu atoms that have paired with an oxygen atom remain on the sample, leading to a stoichiometric EuO film whose growth rate is controlled by the adsorption of the oxygen gas. Furthermore, the europium overpressure inhibits the formation of $Eu_2O_3$.

The integration of a magnetic insulator, such as EuO, with semiconductors is important for spintronic devices since semiconductors are the mainstay of the current electronics industry. While there has been great progress in the MBE growth of $EuO_{1-\delta}$ on silicon and GaN [6,12], the growth of stoichiometric EuO on a semiconductor has yet to be achieved. We are particularly interested in EuO/GaAs because the GaAs is well suited for optical probes of the spin polarization and spin dynamics [13-15]. However, the direct growth of stoichiometric EuO on GaAs presents significant challenges. It is well known that elemental rare-earths grown on GaAs(001) result in highly reacted interfacial phases [16]. In addition, the elevated substrate temperatures required for stoichiometric growth will enhance the interface reaction and also promote interdiffusion. This suggests that a diffusion barrier will be necessary for the integration of stoichiometric EuO with GaAs, similar to the approach utilized for integrating epitaxial oxides onto silicon [12, 17-19].

In this Letter, we report the epitaxial growth of EuO films on GaAs(001) in the adsorption-controlled regime using an MgO diffusion barrier. EuO is deposited on yttrium-stabilized cubic zirconia (YSZ) and MgO substrates to demonstrate high quality, adsorption-controlled growth. When deposited directly onto GaAs, EuO exhibits poor crystalline structure and weak or no ferromagnetic behavior. Therefore, we employ MgO diffusion barriers on GaAs(001) and subsequently deposit EuO overlayers in the adsorption-controlled regime. These samples exhibit high quality single-crystal structure and possess good magnetic properties including a Curie temperature ($T_c$) of 69 K (equal to bulk $T_c$ [3]), hysteresis loops with substantial remanent magnetization, and a large magneto-optic Kerr rotation of 0.57°.

Samples are grown by MBE in an ultrahigh vacuum chamber with a base pressure of $1\times10^{-10}$ torr. For all samples, pure Eu metal (99.99%) is evaporated from a thermal effusion cell at a rate of 7.4-7.8 Å/min. For all EuO depositions, the samples are held at 450 C for the re-evaporation of Eu. Growths are initiated with Eu flux, followed by molecular oxygen gas (99.994% pure) that is leaked into the chamber to a stable pressure of $1.0\times10^{-8}$ torr. For all samples, the growth time is 30 minutes and is terminated by closing the oxygen leak valve and then closing the Eu shutter within 30 s. Samples are capped with 3 nm MgO from an electron beam source to protect the EuO from further oxidation. For growth on GaAs, GaAs(001) substrates with GaAs buffer layers are prepared by III-V MBE and capped with As. After transferring in air to the EuO/MgO MBE system, the As cap is desorbed to yield a 2×4 GaAs(001) surface. An atomic force microscopy (AFM) profile scan of a sample grown under these conditions gives a film thickness of 5.54 ± .07 nm, for which a saturation magnetization value of 6.93 ± 0.26 $\mu_B$/Eu atom is measured by vibrating sample magnetometry at 5 K (not shown). Within the error, this is nearly identical to the theoretical value of 7 $\mu_B$/Eu atom for EuO.

The epitaxy of EuO on YSZ(001) (Y$_2$O$_3$:ZrO 8% mol) serves as a reference for the adsorption-controlled growth due to the excellent lattice match of 0.3% ($a_{EuO}$ = 5.140 Å, $a_{YSZ}$ = 5.125 Å) [11]. Fig. 1a and Fig. 1b show typical reflection high energy electron diffraction (RHEED) patterns for EuO/YSZ(001) along the [100] and [110] directions, respectively. RHEED oscillations (not shown) are seen up to 8 monolayers (ML) independent of the oxygen partial pressure, indicating that the initial growth is substrate assisted in agreement with previous reports [11]. Samples are characterized by longitudinal magneto-optic Kerr effect (MOKE) (635 nm, 150 $\mu$W) in an optical flow cryostat with an angle of incidence of 45° and $p$-polarized incident beam. The inset of Fig. 1c shows a typical MOKE loop for EuO/YSZ measured at 5 K. The coercive field ($H_c$) is 202 Oe, the ratio of the remanent magnetization to saturation magnetization ($M_R$/$M_S$) is 0.69, and the measured Kerr rotation is 0.65°. The remanent Kerr rotation as a function of temperature (Fig. 1c) indicates a $T_c$ of 69 K, in agreement with the bulk $T_c$. These magnetic properties indicate the growth of high quality EuO.

Following the growth on YSZ(001), we deposit EuO on both As- and Ga-terminated surfaces of GaAs(001) at 450 °C. In all cases, RHEED patterns are nonexistent and MOKE characterizations show either weak or no ferromagnetic behavior. The high temperature likely aides in the out-diffusion of As and the formation of reactive phases. A possible method to suppress the interdiffusion and interface reaction is to employ a thin MgO diffusion barrier (i.e. EuO/MgO/GaAs).

To systematically develop the MgO diffusion barrier, we grow EuO on an MgO(001) substrate because the large lattice mismatch of 22.5% ($a_{EuO}$ = 5.140 Å, $a_{MgO}$ = 4.212 Å) makes the epitaxial growth non-trivial and magnetic quality uncertain. Prior to growth, the MgO substrates are annealed at 600 °C in an oxygen environment ($1 \times 10^{-7}$ torr), followed by a 5 nm MgO buffer layer deposited at 350 °C [20] in an oxygen environment ($8 \times 10^{-8}$ torr). The RHEED pattern of the MgO buffer layer is shown in Fig. 1d. At the onset of EuO growth, the RHEED pattern fades, with only the central peak remaining. After approximately 2 nm, the pattern reappears (Fig. 1d), and the side diffraction streaks are shifted by 22% (compared to MgO), indicating high quality epitaxy and cube-on-cube growth. Longitudinal MOKE loops (Fig. 1f) taken at 5 K indicate excellent magnetic properties with an $H_c$ of 58 Oe and $M_R/M_S$ of 0.97.

Since it is well known that the epitaxy of MgO on semiconductors is highly temperature dependent [17, 21-22], we next investigate the growth of MgO diffusion barriers on GaAs(001) and optimize the MgO quality by systematically varying the growth temperature. Starting with a 2×4 GaAs(001) surface (Fig. 2a and 2b), 2 nm MgO films are grown in an oxygen environment ($8 \times 10^{-8}$ torr) at substrate temperatures ranging from 50 C to 420 C (the limit for 2×4 reconstruction). Because the adsorption-controlled growth of EuO requires a substrate temperature of 450 °C, the MgO films are subsequently annealed for 30 minutes at 450 C. After post annealing, samples are removed for *ex situ* AFM. Figures 2g, 2h, and 2i show AFM scans of MgO films grown at 50 C, 300 C, and 400 C, respectively. As summarized in Fig. 2j, AFM scans yield RMS roughness values between 0.540 nm and 0.198 nm, with the optimal growth of MgO on GaAs at 300 C. The corresponding RHEED patterns of the optimized MgO are shown in Fig. 2b and 2c for [100] and [110], respectively, and indicate good single-crystal structure. Lastly, adsorption-controlled EuO is deposited on MgO/GaAs(001) at 450 C with an optimized MgO diffusion barrier. RHEED patterns, as shown in Fig 3e and 3f along [100] and [110], respectively, indicate that the growth of EuO is single crystal and cube-on-cube.

Fig. 3a shows longitudinal MOKE hysteresis loops taken at 5 K, 40 K, and 65 K. At 5 K, the $H_c$ is 102 Oe, $M_R/M_S$ is 0.44, and the saturation Kerr rotation is 0.57°. With increasing temperature, both $H_c$ and the Kerr rotation decrease monotononically as expected for magnetic thin film behavior. Figure 3b shows a detailed measurement of the remanent Kerr rotation as a function of temperature. The measured Curie temperature, $T_c$ = 69 K, agrees with the bulk value. Comparing these properties to the EuO films on YSZ(001) and MgO(001) substrates indicates that the magnetic properties of EuO/MgO/GaAs(001) are weaker but still very good.

In conclusion, we have grown EuO on YSZ(001), MgO(001), and GaAs(001) in the adsorption-controlled regime. For growth on GaAs(001), an MgO diffusion barrier is employed to suppress the interface reaction and interdiffusion between the EuO film and GaAs substrate. All films exhibit a $T_c$ of 69 K, large MOKE signals, and relatively square hysteresis loops. The growth of single-crystal, stoichiometric EuO films on GaAs enables alternative approaches for injecting, detecting, and manipulating spin in GaAs.

We acknowledge the support of NSF, ONR, and CNN/DMEA. We would also like to thank D. Gui, and U. Mohideen for technical assistance.

**Figure Captions**

**Fig. 1.** (a), (b) EuO RHEED patterns on YSZ(001) along [100] and [110], respectively. (c) Temperature dependence of the Kerr rotation at saturation for EuO/YSZ(001). Inset: longitudinal MOKE hysteresis loop at 5 K. (d) RHEED pattern of 5 nm MgO buffer layer on MgO(001). (e) RHEED pattern of EuO/MgO(5 nm)/MgO(001) along [100]. (f) MOKE hysteresis loop of EuO/MgO(5 nm)/MgO(001).

**Fig. 2.** (a) – (f) RHEED patterns of the GaAs substrate with 2×4 reconstruction and subsequent growths of 2 nm MgO grown at 300 C and EuO grown at 450 C. The left column is in the [100] direction and the right column is in the [1$\bar{1}$0] direction. (g) – (i) AFM measured at RT on 2 nm MgO grown at 50 C, 300 C and 400 C, respectively, on 2×4 reconstructed GaAs. (j) RMS roughness of 2 nm MgO on GaAs displayed as a function of growth temperature.

**Fig. 3.** Magnetization characterization by MOKE on EuO /MgO(2 nm)/GaAs (001). (a) Longitudinal MOKE hysteresis loops taken at 5 K, 40 K, and 65 K. (b) Temperature dependence of the Kerr rotation at saturation.

Fig. 1

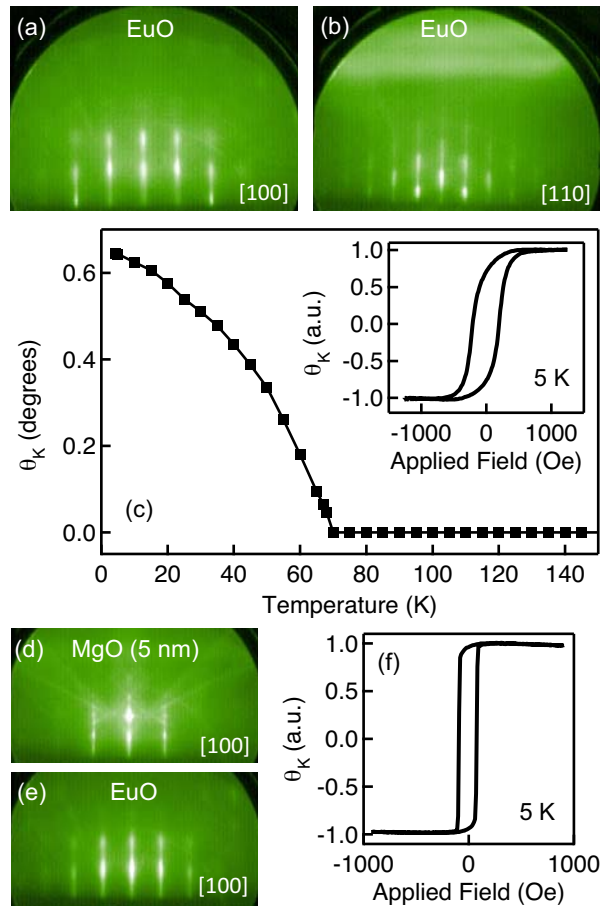

Fig. 2

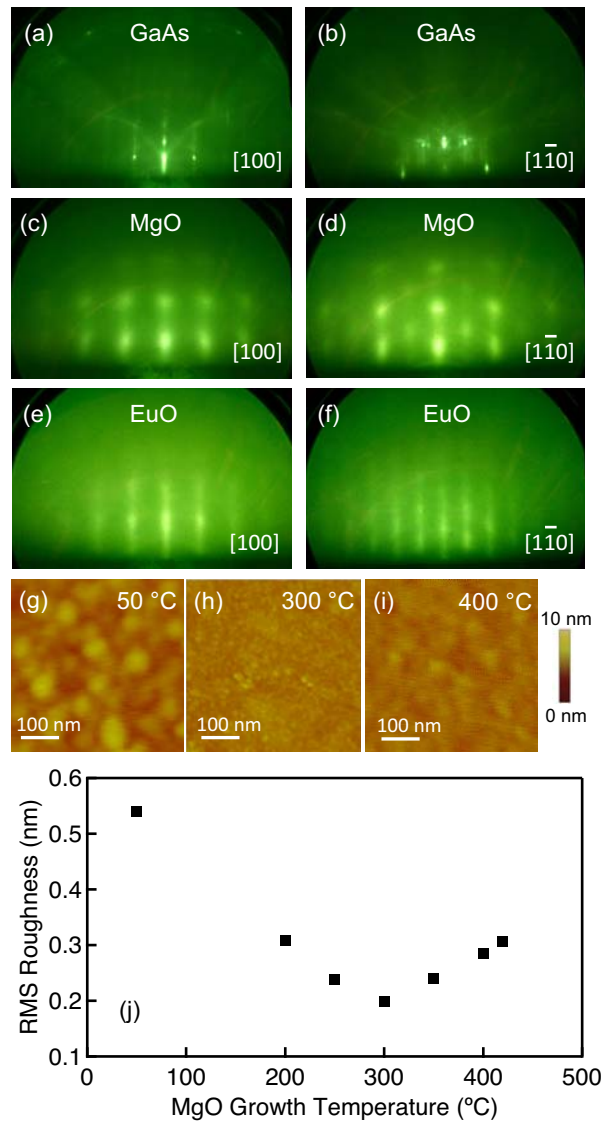

Fig. 3

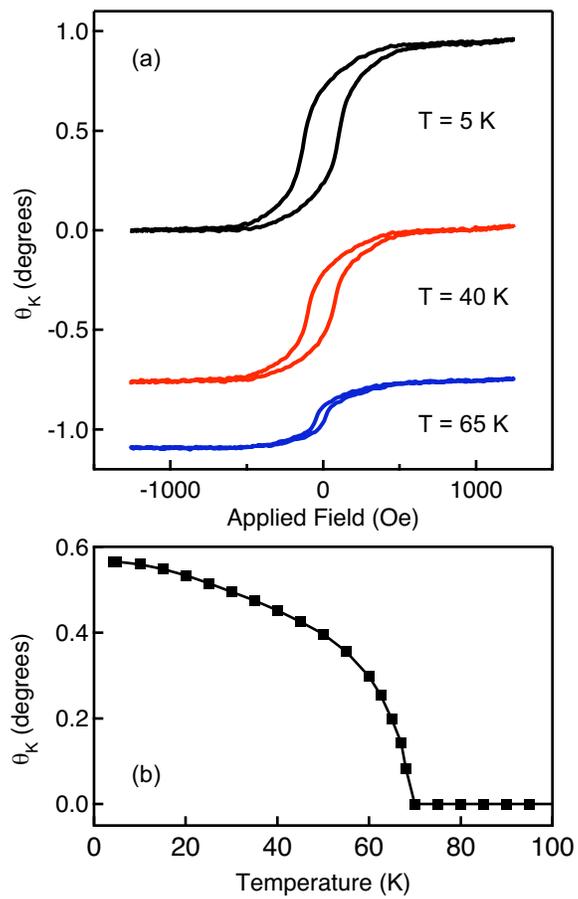